\documentclass[aps,pre,nofootinbib,amsfonts,superscriptaddress,longbibliography,showkeys]{revtex4-1}
\usepackage{hyperref}
\usepackage{graphicx,bm,amsmath,color}
\usepackage[latin1]{inputenc}
\usepackage[english]{babel}
\usepackage{bm}
\usepackage[T1]{fontenc}
\usepackage{epstopdf}
\usepackage{booktabs}
\usepackage{placeins}
\usepackage{multirow}
\newcommand{\ba}{\begin{array}}
\newcommand{\ea}{\end{array}}

\begin{document}

\title{Collective social behavior in a crowd controlled game}

\author{Alberto Aleta}
\email{albertoaleta@gmail.com}
\affiliation{Institute for Biocomputation and Physics of Complex Systems (BIFI), Universidad de Zaragoza, 50018 Zaragoza, Spain}
\affiliation{Department of Theoretical Physics,
Universidad de Zaragoza, Zaragoza 50009, Spain}
\author{Yamir Moreno}
\email{yamir.moreno@gmail.com}
\affiliation{Institute for Biocomputation and Physics of Complex Systems (BIFI), Universidad de Zaragoza, 50018 Zaragoza, Spain}
\affiliation{Department of Theoretical Physics,
Universidad de Zaragoza, Zaragoza 50009, Spain}
\affiliation{ISI Foundation, Turin, Italy}


\begin{abstract}
Despite many efforts, the behavior of a crowd is not fully understood. The advent of modern communication media has made it an even more challenging problem, as crowd dynamics could be driven by both human-to-human and human-technology interactions. Here, we study the dynamics of a crowd controlled game (Twitch Plays Pok\'emon), in which nearly a million players participated during more than two weeks. We dissect the temporal evolution of the system dynamics along the two distinct phases that characterized the game. We find that players who do not follow the crowd average behavior are key to succeed in the game. The latter finding can be well explained by an n-$th$ order Markov model that reproduces the observed behavior. Secondly, we analyze a phase of the game in which players were able to decide between two different modes of playing, mimicking a voting system. Our results suggest that under some conditions, the collective dynamics can be better regarded as a swarm-like behavior instead of a crowd. Finally, we discuss our findings in the light of the social identity theory, which appears to describe well the observed dynamics.  
\end{abstract}

\maketitle

Collective phenomena have been the subject of intense research in psychology and sociology since the XIX century. There are several ways in which humans gather to perform collective actions, although observations suggest that most of them require some sort of diminution of self-identity \cite{abrams2001collective}. One of the first attempts to address this subject was Le Bon's theory on the psychology of crowds in which he argued that when people are part of a crowd they lose their individual consciousness and become more primitive and emotional thanks to the anonymity provided by the group \cite{LeBon1895}. In the following decades, theories of crowd behavior such as the convergence theory, the emergent norm theory or the social identity theory emerged. These theories shifted away from Le Bon's original ideas, introducing rationality, collective norms and social identities as building blocks of the crowd \cite{reicher2001psychology} \cite{LaMacchia2016}. 

The classical view of crowds as an irrational horde led researchers to focus on the study of crowds as something inherently violent, and thus, to seek for a better understanding and prediction of violence eruption, or at least, to develop some strategies to handle them \cite{reicher2004integrated}. However, the information era has created a new kind of crowd, as it is no longer necessary to be in the same place to communicate and take part of collective actions. Indeed, open source and ``wiki'' initiatives, as well as crowdsourcing and crowdworking, are some examples of how crowds can collaborate online in order to achieve a particular objective \cite{Kozinets2008} \cite{vonAhn2008Sep}. Although this offers a plethora of opportunities, caution has to be taken because, as research on the psychology of crowds has shown, the group is not just the simple addition of individuals \cite{baumeister2016}. For example, it has been observed that the group performance can be less than the sum of the individual performances if they had acted separately \cite{Latane1981}, which represents a challenge if one wants to use crowds as a working force.

To be able to unlock the potential of collective intelligence, a deeper understanding of the functioning of these systems is needed \cite{Quinn2011May}. Examples of scenarios that can benefit from further insights into crowd behavior include new ways to reach group decisions, such as voting, consensus making or opinion averaging, as well as finding the best strategies to motivate the crowd to perform some task \cite{malone2010collective}. Nevertheless, as arbitrary tasks usually are not intrinsically enjoyable, to be able to systematically execute crowdsourcing jobs, some sort of financial compensation is used \cite{mason2010financial}. This, however, implies dealing with new challenges, since many experiments have demonstrated that financial incentives might undermine the intrinsic motivation of workers or encourage them to only seek for the results that are being measured, either by focusing only on them or by free-riding \cite{prendergast1999provision} \cite{heyman2004effort} \cite{gneezy2000pay}. A relevant case is given by platforms such as Amazon's Mechanical Turk, that allow organizations to pay workers that perform micro-tasks for them, and that have already given rise to interesting questions about the future of crowd work \cite{Kittur2013}. In particular, its validity to be used for crowdsourcing behavioral research has been recently called into question \cite{Peer2017}.

Notwithstanding the previous observations, it is possible to find tasks that are intrinsically enjoyable by the crowd due to their motivational nature, which is ultimately independent of the reward \cite{gneezy2000pay}. This is one of the basis of online citizen science. In these projects, volunteers contribute to analyze and interpret large datasets which are later used to solve scientific problems \cite{Cox2015Jul}. To increase the motivation of the volunteers, some of these projects are shaped as computer games \cite{vonAhn2006Jun}. Examples range from the study of protein folding \cite{Khatib2011Nov} to annotating people within social networks \cite{Bernstein2009Oct} or identifying the presence of cropland \cite{Salk2016Apr}.

It is thus clear that to harness the full potential of crowds in the new era, we need a deeper understanding of the mechanisms that drive and govern the dynamics of these complex systems. To this aim, in this paper we study an event that took place in February 2014 known as Twitch Plays Pok\'emon (TPP). During this event, players were allowed to control simultaneously the same character of a Pok\'emon game without any kind of central authority. This resulted in a completely crowd controlled process in which thousands of users played simultaneously for 17 days, with more than a million different players \cite{Records2014Nov}. TPP is specially interesting because it represents an out of the lab social experiment that became extremely successful based only on its intrinsic enjoyment and, given that it was run without any scientific purpose in mind, 
it represents a natural, unbiased (i.e., not artificially driven) opportunity to study the evolution and organization of crowds.

\section{Description of the event\label{section:description}}

On February 12, 2014, an anonymous developer started to broadcast a game of Pok\'emon Red on the streaming platform Twitch \cite{Records2014Nov}. Pok\'emon Red was the first installment of the Pok\'emon series, which is the most successful role playing game (RPG) franchise of all time \cite{rpgSales}. The purpose of the game was to capture and train creatures known as Pok\'emons in order to win increasingly difficult battles based on classical turn based combats. However, as Pok\'emon Go showed in the summer of 2016, the power of the Pok\'emon franchise goes beyond the classical RPG games and is still able to attract millions of players \cite{Althoff2016Dec}. 

On the other hand, Twitch is an online service for watching and streaming digital video broadcast. Its content is mainly related to video games: from e-sports competitions to professional players games or simply popular individuals who tend to gather large audiences to watch them play, commonly known as streamers. Due to the live nature of the streaming and the presence of a chat window where viewers can interact among each other and with the streamer, the relationship between the media creator and the consumer is much more direct than in traditional media \cite{Sjoblom2017Oct}. Back in February 2014 Twitch was the 4th largest source of peak internet traffic in the US \cite{twitchTraffic} and nowadays, with over 100 million unique users, it has become the home of the largest gaming community in history \cite{Churchill2016Oct}.

The element that distinguished this stream from the rest was that the streamer did not play the game. Instead, he set up a bot in the chat window that accepted some predefined commands and forwarded them to the input system of the video game. Thus, anyone could join the stream and control the character by just writing one of those actions in the chat. Although all actions were sent to the video game sequentially, it could only perform one at a time. As a consequence, all commands that arrived while the character was performing a given action (which takes less than a second) did not have any effect. Thus, it was a completely crowd controlled game without any central authority or coordination system in place. This was not a multiplayer game, this was something different, something new \cite{wellPlayed}.

\begin{figure}%
\begin{center}
\includegraphics[width=\linewidth]{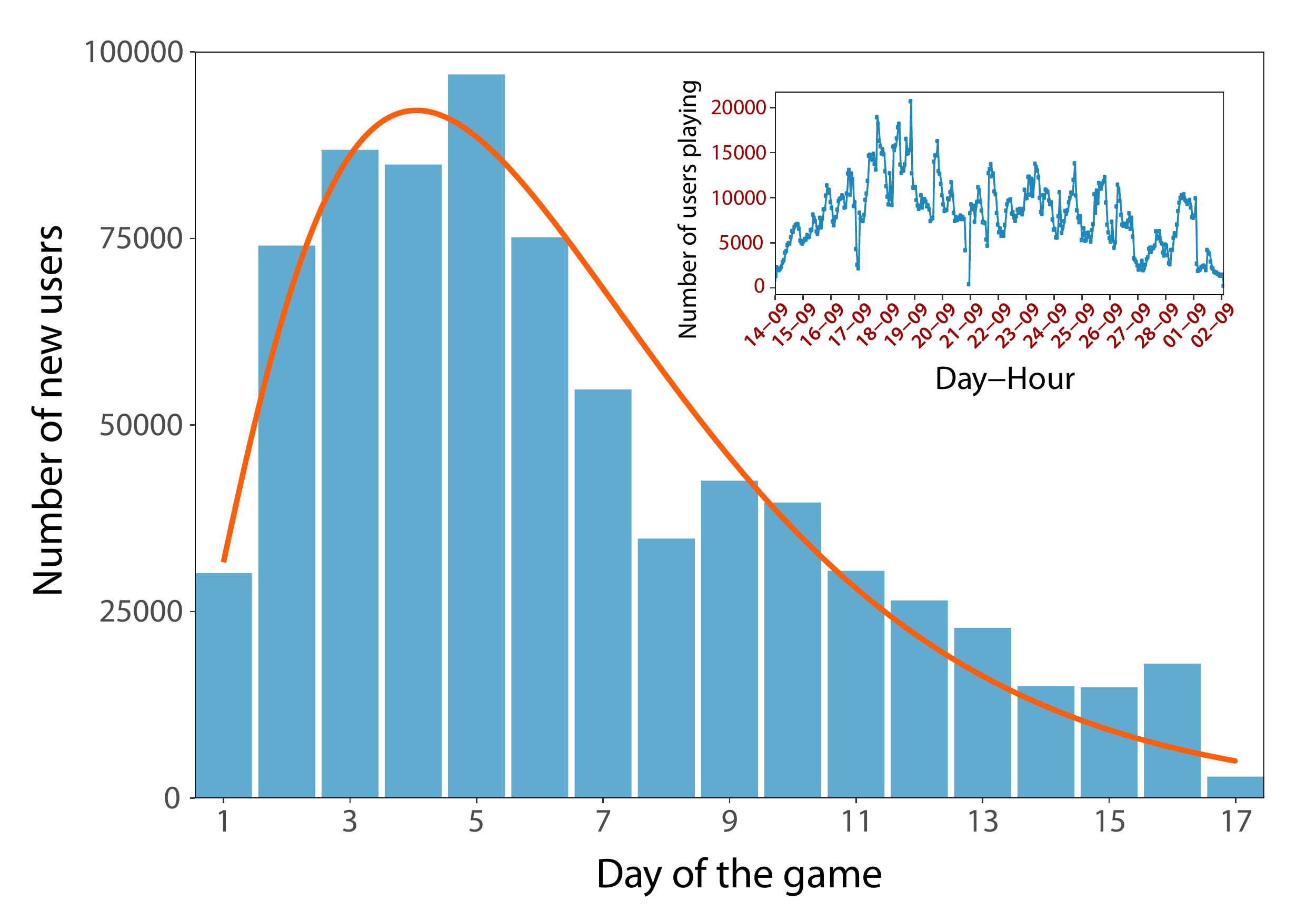}%
\end{center}
\caption{Number of new users per day. The histogram is fitted to a gamma distribution of parameters $\alpha=2.66$ and $\beta=0.41$. Note that this reflects those users who inputted at least one command, not the number of viewers. In the inset we plot the total number of users playing per hour, regardless on whether they were new players or not. All timestamps are GMT+1.
}%
\label{fig:new_users}%
\end{figure}

Due to its novelty, during the first day the game was mainly unknown with only a few tens of viewers/players and as a consequence little is known about the game events of that day \cite{timelineTPP}. However, on the second day it started to gain viewers and quickly went viral, see figure \ref{fig:new_users}. Indeed, it ramped up from 25,000 new players on day 1 (note that the time was recorded starting from day 0 and thus day 1 in game time actually refers to the second day on real time) to almost 75,000 on day 2 and an already stable base of nearly 10,000 continuous players. Even though there was a clear decay on the number of new users after day 5, the event was able to retain a large user base for over two weeks. This huge number of users imposed a challenge on the technical capabilities of the system, which translated in a delay of between 20 and 30 seconds between the stream and the chat window. That is, users had to send their commands based on where the player was 30 seconds ago.

Although simple in comparison to modern video games, Pok\'emon Red is a complex game which can not be progressed at random. In fact, a single player needs, on average, 26 hours to finish the game \cite{timePokemon}. Nevertheless, only 7 commands are needed to complete the game. There are 4 movement commands (\emph{up}, \emph{right}, \emph{down} and \emph{left}), 2 actions commands (\emph{a} and \emph{b}, usually used as \emph{accept} and \emph{back/cancel}) and 1 system button (\emph{start} which opens the game's menu). As a consequence the gameplay is simple. The character is moved around the map using the four movement commands. If you encounter a wild Pok\'emon you will have to fight it with the possibility of capturing it. Then, you will have to face the Pok\'emons of trainers controlled by the machine in order to obtain the 8 medals needed to finish the game. The combats are all turn-based so that time is not an important factor. In each turn of a combat the player has to decide which action to take for which the movement buttons along with \emph{a} and \emph{b} are used. Once the 8 medals have been collected there is a final encounter after which the game is finished. This gameplay, however, was much more complex during TPP due to the huge number of players sending commands at the same time and the lag present in the system. This lag was a result of the technical difficulties of allowing so many individuals to interact simultaneously in a platform that was not designed to do so. As a consequence, there was a delay of around 20-30 seconds between the chat input and the video, so that the image would actually reflect were the player was 20-30 seconds ago.

A remarkable aspect of the event is that actions that would usually go unnoticed, such as nicknaming the Pok\'emons, yielded unexpected outcomes due to the messy nature of the gameplay. The community embraced these outcomes and created a whole narrative around them in the form of jokes, fan art and even a religion-like movement based on the judeo-christian tradition \cite{religion2015} both in the own chat window and in related media such as Reddit. Although these characteristics of the game are outside of the scope of the paper, we believe that it would be interesting to address them as an example of the evolution of naming conventions and narrative consensus \cite{Centola2015Feb}.

Even if it was at a slower peace, progress was made. Probably the first thing that comes to ones mind when thinking on how progressing was possible is the famous experiment by Francis Galton in which he asked a crowd to guess the weight of an ox. He found that the average of all estimates of the crowd was just 0.8\% higher than the real weight \cite{galton1907vox}. Indeed, when lots of users are playing, the extreme answers would cancel each other and the character would tend to move towards the most common command sent by the crowd. Note, however, that as they were not voting, actions deviating from the mean could also be performed by pure chance. In general, this did not have great effects but as we will see in section \ref{subsection:ledge} there were certain parts of the game where this was extremely relevant. 

\begin{figure}
\begin{center}
\includegraphics[width=\linewidth]{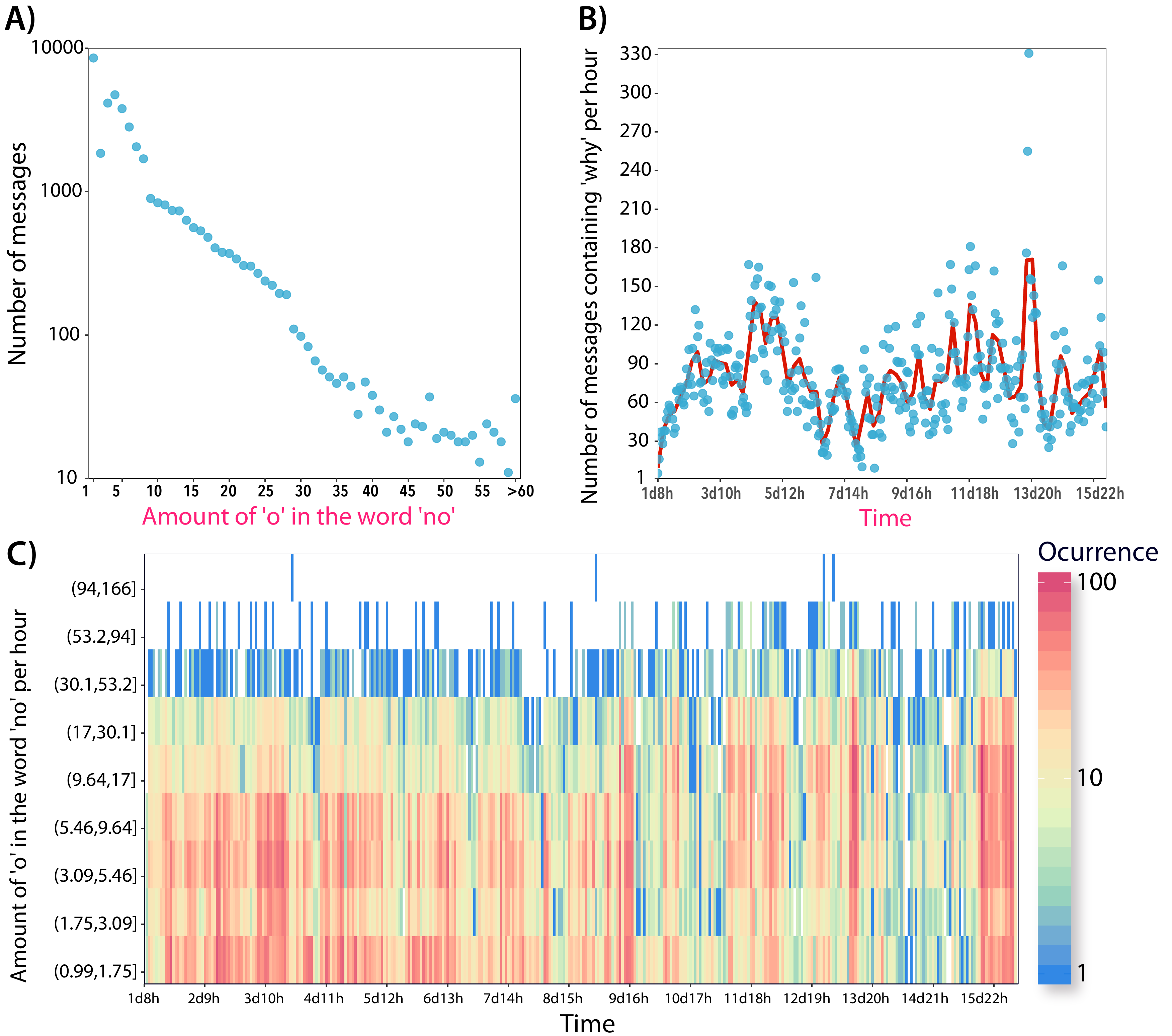}%
\end{center}
\caption{Measures of frustration. A) Players expressed their frustration by adding more times the letter ``o'' when they wanted to say ``no''. Interestingly, the relationship is not linear as the word ``noo'' tends to appear less than ``nooo'' or ``noooo'', which indicates that when players were frustrated they overexpress it. B) Number of messages containing the word ``why'' per hour. This indicates that many players did not understand the movements of the crowd, which probably made them feel frustrated. C) Time evolution of the amount of letters ``o'' in the word ``no''. Frustration was clearly something present throughout the whole event. }%
\label{fig:frustration}%
\end{figure}

It is worth stressing that, to form a classical wise crowd, some important elements are needed, such as independence \cite{Surowiecki2004}. That is, the answer of each individual should not be influenced by other people in the crowd. In our case, this was not true, as the character was continuously moving. Indeed, the big difference of this crowd event to others is that opinions had effect in real time, and hence, people could see the tendency of the crowd and change its behavior accordingly. Theoretical \cite{couzin2005effective} and empirical studies \cite{dyer2008consensus} have shown that a minority of informed individuals can lead a na\"ive group of animals or humans to a given target in the absence of direct communication. Even in the case of conflict in the group, the time taken to reach the target is not increased significantly \cite{dyer2008consensus} which would explain why it only took the crowd 10 times more to finish the game than the average person. Although this ration may seem high, as we shall see later, the crowd got stuck in some parts of the game for over a day, increasing the time to finish. However, if those parts were excluded, the game progress can be considered to be remarkably fast, despite the messy nature of the gameplay.

As a matter of fact, the movement of the character on the map can be probably better described as a swarm rather than as a crowd. Classical collective intelligence, such as the opinions of a crowd obtained via polls or surveys, has the particularity stated previously of independence and in addition to being asynchronous. Even more, it has been shown that when users can influence each other but still in an asynchronous way, the group decisions are distorted by social biasing effects \cite{Muchnik2013Aug}. Recently, it has been proposed that the use of structures similar to natural swarms can correct some of these problems \cite{Rosenberg2015Sep}. Indeed, by allowing users to participate in decision making processes in real time with a feedback about what the rest is doing, in some sort of human swarm, it is possible to explore more efficiently the decision space and reach more accurate predictions than with simple majority voting \cite{Rosenberg2016Oct}. Admittedly, it has recently been suggested that online crowds might be better described as swarms as something in-between crowds and networks \cite{Lee2017jan}.

\begin{figure}[t]
\centering
\includegraphics[width=14cm,height=6.4cm]{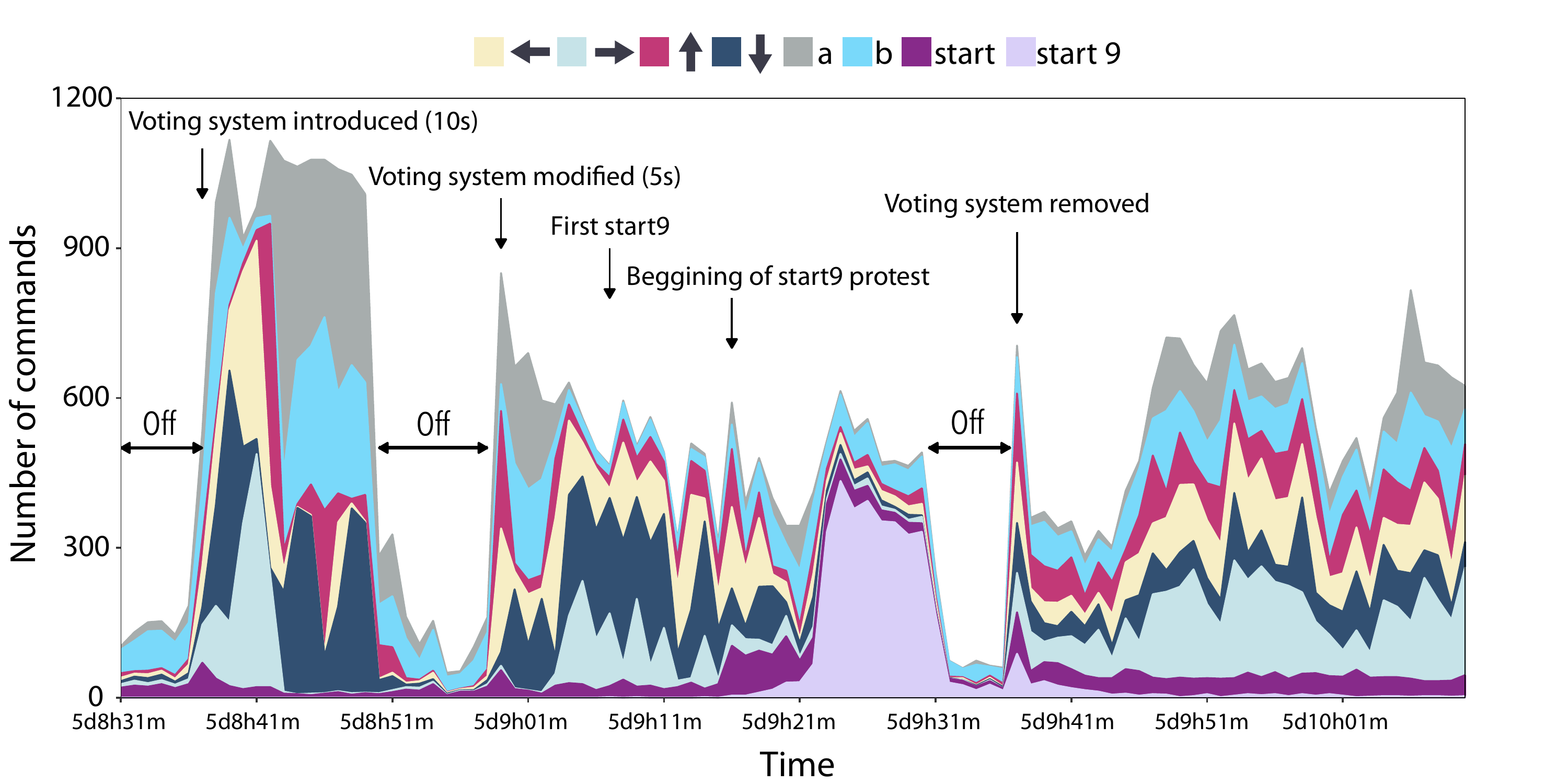}%
\caption{Command distribution after the first introduction of the voting system. Once the system was back online votes would tally up over a period of 10 seconds. After 15 minutes the system was brought down to reduce this time to 5 seconds. This, however, did not please the crowd and started to protest. The first $start9$ was sent at 5d9h8m but went almost unnoticed. Few minutes after it was sent again but this time it got the attention of the crowd. In barely  3 minutes it went from 4 $start9$ per minute to over 300, which stalled the game for over 8 minutes. The developer brought down the system again and removed the voting system, introducing the anarchy/democracy system a few hours later.
}%
\label{fig:start9}%
\end{figure}

Even though the characteristics described so far already make this event very interesting, on the sixth day the rules were slightly changed, which made the dynamics even richer. After the swarm had been stuck in a movement based puzzle for almost 24 hours, the developer took down the stream to change the code. Fifteen minutes later the stream was back online but this time commands were not executed right away. Instead, they were added up and every 10 seconds the most voted command was executed. In addition, it was possible to use compound commands made of up to 9 simple commands such as $a2$ or $aleftright$ which would correspond to executing $a$ twice or $a$, $left$ and $right$ respectively. Thus, the swarm became a crowd with a majority rule to decide which action to take. As it waited 10 seconds between each command, progress was slow and, twenty minutes after, that time was reduced to 5 seconds. However, the crowd did not like this system and started to protest by sending \emph{start9} which would open and close the menu repeatedly impeding any movement. This riot, as it was called, lasted for 8 minutes (figure \ref{fig:start9}), moment when the developer removed the voting system. However, two hours later the system was modified again. Two new commands were added: \emph{democracy} and \emph{anarchy}, which controlled some sort of tug-of-war voting system over which rules to use. If the fraction of people voting for democracy went over a given threshold, the game would start to tally up votes about which action to take next. If not, the game would be played using the old rules. This system split the community into ``democrats'' and ``anarchists'' who would fight for taking control of the game. Therefore, the system would change between a crowd-like movement and a swarm-like movement purely based on its own group interactions. This situation will be analyzed in section \ref{subsection:anarchy}.

To finish this section we should discuss the reasons why this game might have attracted so many people. The game was disordered, progress was slower than if played individually, and often really bad actions were taken (such as mistakenly releasing some of the strongest Pok\'emons) which led to frustration. Indeed, by looking at the chat logs it is possible to measure the frustration the players felt during the event, figure \ref{fig:frustration}. Although usually frustration has a negative connotation, in the context of games it has been observed that frustration and stress can be pleasurable as they motivate players to overcome new challenges \cite{Nylund2015}.  Actually, there is a whole game genre known as ``masocore'' (a portmanteau of masochism and hardcore) which consists of games with extremely challenging gameplay built with the only purpose of generating frustration on the players \cite{masocore}. Similarly, there are games which might be simpler but that have really difficult controls and strange physics, such as QWOP, Surgeon Simulator or Octodad, which are also built with the sole aim of generating frustration \cite{bustedPhys}. Thus, the mistakes performed by the crowd might have not been something dissatisfactory but completely the opposite, they might have been the reason why this event was so successful.

\section{Results}

We divide the results section into two subsections. In the first one, we will study the event know as ``the ledge''. This is the name the crowd gave to an area that is extremely easy to finish for a single player but that represented a hard challenge for them, so that we will be able to explore the collective behavior of the crowd at short timescales. Then, on the second subsection we will analyze the political movements that evolved within the game, giving us information about the behavior of the crowd at longer timescales.

\subsection{The Ledge\label{subsection:ledge}}

\begin{figure*}%
\begin{center}
\includegraphics[width=17cm]{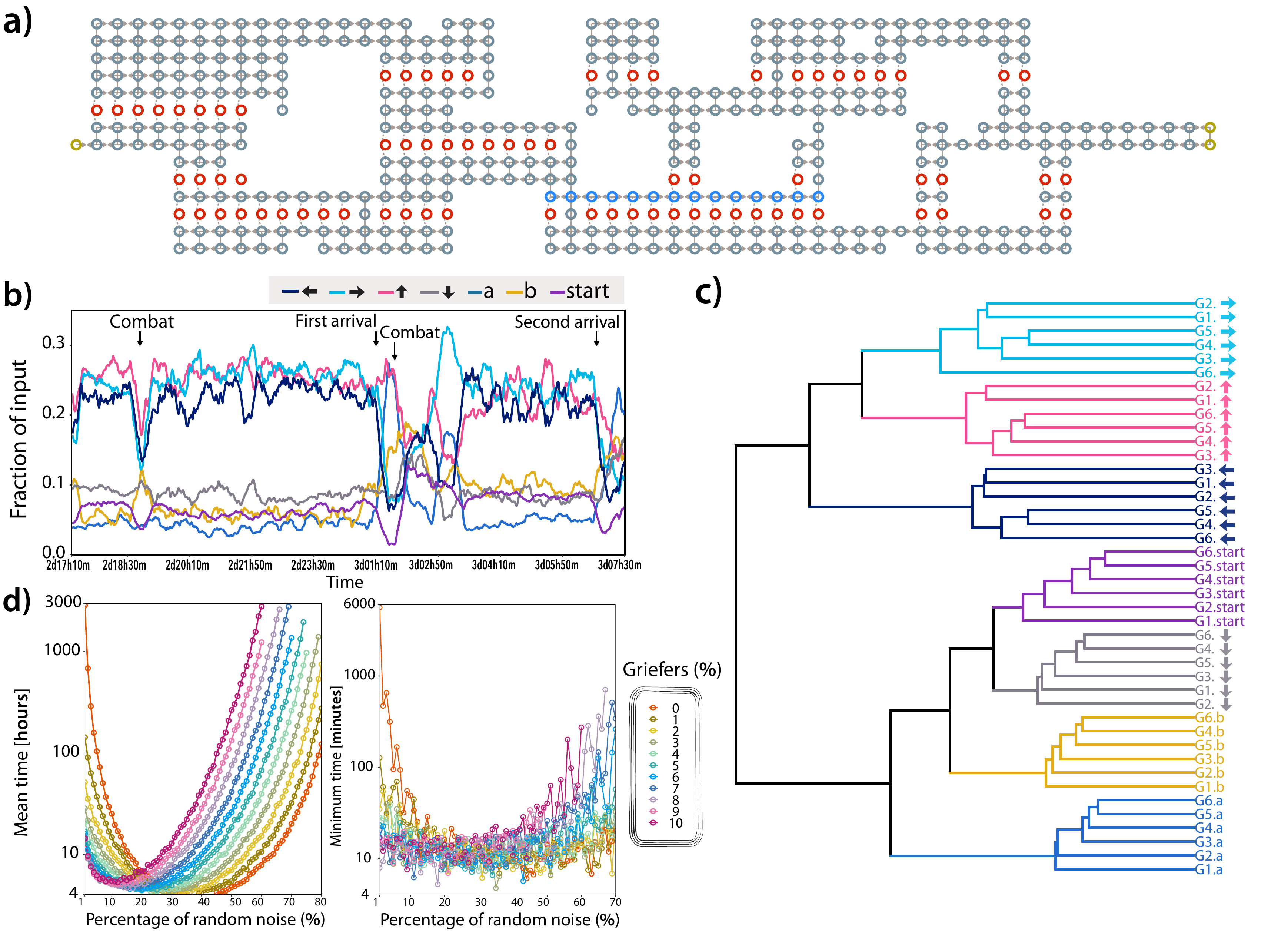}%
\end{center}
\caption{Study of the Ledge event. a) Network representation of the area. It is possible to go from a node to the ones surrounding it using the commands \emph{up}, \emph{right}, \emph{down} and \emph{left}. The only exception are the red nodes L which represent ledges. If the character tries to step on one of those nodes it will be automatically sent to the node right below it, characteristic that is represented by the curved links connecting nodes above and below ledges. Yellow nodes mark the entrance and exit of the area and blue nodes highlight one of the most difficult parts of the path. b) Time evolution of the fraction of commands sent each minute. The time series has been smoothed using moving averages. c) Hierarchical clustering of the time series of each group of users (see main text for details). d) Left: Mean time needed to exit the area according to our simulations as a function of the fraction of griefers in the system and the noise in it. Right: minimum time needed to exit the area, note that the y axis is given in minutes instead of hours.
}%
\label{fig:ledge}%
\end{figure*}

On the third day of the game, the character arrived to the area depicted in figure \ref{fig:ledge}a (note that the democracy/anarchy system had not been introduced yet). Each node of the graph represents a tile of the game $-$as the map of the game is discrete, this network representation is not an approximation, but the exact shape of the area. The character starts on the yellow node on the left part of the network and has to exit through the right part, event that we will define as getting to one of the yellow nodes on the right. The path is simple for an average player but it represents a challenge for the swarm due to the presence of the red nodes. These nodes represent ledges which can only be traversed going downwards, effectively working as a filter that allows flux only downwards. Thus, one good step will not cancel a bad step, as the character would be trapped down the ledge and will have to find a different path to go up again. For this reason, this particular region is highly vulnerable to actions deviating from the norm, either caused by mistake or performed intentionally by griefers, i.e., individuals whose only purpose is to annoy other players and who do so by using the mechanisms provided by the game itself \cite{Kirman2012} \cite{paul2015enjoyment} (note that in social contexts these individuals are usually called trolls \cite{Buckels2014}). Indeed, there are paths (see blue nodes in figure \ref{fig:ledge}a) where only the command \emph{right} is needed and which are next to a ledge, so that the command \emph{down} which is not needed at all will force the swarm to go back and start the path again. Additionally, the existence of the lag described in section \ref{section:description} made this task even more difficult.

In figure \ref{fig:ledge}b we show the time evolution of the amount of messages containing each command (the values have been normalized to the total number of commands sent each minute) since the beginning of this part until they finally exited. First, we notice that it took the swarm over 15 hours to finish an area that can be completed by an optimal walk in less than 2 minutes. Then, we can clearly see a pattern from 2d18h30m to the first time they were able to reach the nodes located right after the pink ones, approximately 3d01h10m: when the number of \emph{rights} is high the number of \emph{lefts} is low. This is a signature of the character trying to go through the pink nodes by going right, falling down the ledge, and going left to start over. Once they finally reached the nodes after the pink path (first arrival) they had to fight a trainer controlled by the game, combat which they lost and as a consequence the character was transported outside of the area and they had to enter and start again from the beginning. Again, we can see a similar left-right pattern until they got over that pink path for the second time, which in this case was definitive.

This area is a great case study of the behavior of the swam because the mechanics needed to complete it are very simple (just moving from one point to another), which facilitates the analysis. At the same time, it took the swarm much longer to finish it than what is expected for a single player. To address all these features, we propose a model aimed at mimicing the behavior of the swarm. Specifically, we consider a $n$-th order Markov Chain so that the probability of going from state $x_{m}$ to $x_{m+1}$ depends only on the state $x_{m-n}$, thus accounting for the effect of the lag in the dynamics. There are 467 nodes in the network, which results in 1868 different states as the character can be in each node facing each of the 4 directions. Indeed, to move in a given direction, the character needs one command if it is already heading that direction or two if it is not, one to point to the desired direction and another to move. Finally, we use the different probabilities to go from one node to another as extracted from the behavior of the players in the swarm.

To define these probabilities, we first classify the players in groups according to the total number of commands they sent in this period: G1, users with 1 or 2 commands (46\% of the swarm); G2, 3 or 4 commands (18\%); G3, between 5 and 7 commands (13\%); G4, between 8 and 14 commands (12\%); G5, between 15 and 25 commands (6\%); and G6, more than 25 commands (5\%). Interestingly, the time series of the inputs of each of these groups are very similar (see supplementary material, S1 to S7). Actually, if we remove their labels and cluster the time series using the euclidean distance, we obtain the same number of clusters as different commands are available. Even more, the time series of each of the commands are clustered together, figure \ref{fig:ledge}c. In other words, the behaviors of users with medium and large activities are not only similar to each other, but they are also equivalent to the ones coming from the aggregation of the users who only sent 1 or 2 commands.

In this context we could argue that users with few messages tend to act intuitively as they soon lose interest. According to the social heuristics hypothesis \cite{Rand2014}, fast decisions tend to increase cooperation, which in this case would mean trying to get out of the area as fast as possible. Similarly, experiments have shown that people with prosocial predispositions tend to act that way when they have to make decisions quickly \cite{yamagishi2017response}. Thus, users that send few commands might tend to send the ones that get the character closer to the exit, which would explain why without being aware of it, they behave as those users that tried to progress for longer. However, coordination might not be so desirable in this occasion. The problem with players conforming with the majoritarian direction or mimicking each other is that they will be subject to herding effects \cite{Kameda2011Jan} \cite{Hung2001Dec} which in this particular setting can be catastrophic due to the lag present in the system. Indeed, if we set the probabilities in our model so that the next state in the transition is always the one that gets you closer to the exit but with 25 seconds of delay (that is, the probability of going from state $x_m$ to $x_{m+1}$ is the probability of going from $x_{m-n}$ to the state which follows the optimal path), the system gets stuck in a loop and is never able to reach the exit.

Nevertheless, perfect coordination is extremely difficult to achieve. Thus, to make our model more realistic we consider that each time step there are 100 users with different behaviors introducing commands. In particular, we consider variable quantities of noisy users who play completely at random, griefers who only press down to annoy the rest of the crowd and the herd who always sends the optimal command to get to the exit. The results, figure \ref{fig:ledge}d, show that the addition of noise to the herd breaks the loops and allows the swarm to get to the exit. In particular, for the case with no griefers we find that with 1 percent of users adding noise to the input the mean time needed to finish this part is almost 3000 hours. However, as we increase the noise, time is quickly reduced with an optimal noise level of around 40\% of the swarm. Conversely, the introduction of griefers in the model, as expected, increases the time needed to finish this part in most cases. Interestingly though, for low values of the noise, the addition of griefers can actually be beneficial for the swarm, allowing the completion of this area in times compatible to the observed ones $-$ also note that the minimum time needed is less dependent on the number of noisy individuals as well as on the number of griefers. Indeed, by breaking the herding effect, these players are unintentionally helping the swarm to reach their goal. 

Whether the individuals categorized as ``noise'' were producing it unintentionally or doing it on purpose to disentangle the swarm is something we can not analyze because, unfortunately, the resolution of the chat log in this area is in minutes and not in seconds. We can, however, approximate the fraction of griefers in the system thanks to the special characteristics of this area. Indeed, as most of the time, the command $down$ is not needed $-$on the contrary, it will destroy all progress$-$, we can categorize those players with an abnormal number of $downs$ as griefers. To do so, we take the users that belong to $G6$ (the most active ones) and compare the fraction of their inputs that corresponds to $down$ between each other. We find that 7\% have a behavior that could be categorized as outlier (the fraction of their input corresponding to $down$ is higher than 1.5 times the inter quartile range). More restrictively, for 1\% of the players, the command $down$ represents more than half of their inputs. Both these values are compatible to the observed time according to our model, even more if we take into account that it is more restrictive as we consider that griefers continuously press down. Thus, we conclude that users deviating from the norm, regardless of being griefers, noise or very smart individuals, were the ones that made finishing this part possible.

\subsection{Anarchy vs. Democracy\label{subsection:anarchy}}

As already commented, on the sixth day of the game, the input system was modified. Commands would be tallied up for 10 seconds (later it was modified to 5 seconds) and the most voted one would be sent to the game. Players did not like this system and started to protest by sending the command $start9$ which would open and close the menu repeatedly effectively stalling the game (figure \ref{fig:start9}). As a consequence, the developer went back to the original setting for a while, but changed it once again 2 hours later to introduce the anarchy/democracy system that we will discuss in this section. This system introduced two new commands, $anarchy$ and $democracy$, which controlled the game mode. If the fraction of players sending democracy went over $0.75$ (later modified to $0.80$) the game would enter into democracy mode and commands would be tallied up for 5 seconds. Then, the meter needed to go below $0.25$ (later modified to $0.50$) to enter into anarchy mode again.

The introduction of the voting system was mainly motivated by a puzzle where the crowd had been stuck for over 20 hours with no progress, but it was not the only reason. It was known that later in the game the character would need to get through a special area, the safari zone, where the number of steps one can take is limited, if it goes over that limit the character would be automatically teleported outside of the area. Voting seemed the best way to go through this area because even though progress can be slow, it is only necessary half of the crowd, at most, to be coordinated. However, even with this system, progressing was still complex as it needed to take over the control of the game mode and take into account lag when trying to progress in democracy mode. Indeed, the tug-of-war system was introduced at the middle of day 5, yet the puzzle was not fully completed until the beginning of day 6, over 40 hours after the crowd had originally arrived to the puzzle.

One of the reasons why it took so long to finish it even after the introduction of the system is that it was very difficult to enter into democracy mode. Democracy was only allowed by the crowd when they were right in front of the puzzle and they would go into anarchy mode quickly after finishing it. Similarly, the rest of the game was mainly played under anarchy mode. Interestingly, though, we find that there were more ``democrats'' in the crowd (players who only voted for democracy) than ``anarchists'' (players who only voted for anarchy). Out of nearly 400,000 players who participated in the tug-of-war throughout the game, 54\% were democrats, 28\% anarchists and 18\% voted at least once for both of them. Therefore, the introduction of this new system did not only split the crowd into two different groups with, as we shall see, their own norms and behaviors, but also created non trivial dynamics between them.

The first question that arises is what might have motivated players to join into one group or the other. From a broad perspective, it has been proposed that one of the key ingredients behind video game enjoyment is the continuous perception of one's causal effects on the environment, also known as effectance \cite{White1959}, thanks to their immediate response to player inputs. In contrast, a reduction of control, defined as being able to influence the dynamics according to one's goals, does not automatically lower enjoyment \cite{Klimmt2007dec}. This might explain why some people preferred anarchy. Under its rules, players saw that the game was continuously responding to inputs, even if they were not exactly the ones they sent. On the other hand, with democracy, control was higher at the expense of effectance, as the game would only advance once every 5 seconds. The fact that some people might have preferred this mode is not surprising as it is well known that different people might enjoy different aspects of a game \cite{mekler2014systematic}. In the classical player classification proposed by Bartle \cite{bartle1996} for the context of MUDs (multi-user dungeon, which later evolved into what we now today as MMORPGs - massively multiplayer online role-playing games) he already distinguished four types of players: achievers, who focus on finishing the game (who in our context could be related to democrats); explorers, who focused on interacting with the world (anarchists); socializers, who focused on interacting with other players (those players who focused on making fan art and developing narratives); and killers, whose main motivation was to kill other players (griefers). Similarly, it has been seen in the context of FPSs (first person shooters) that player-death events, i.e., loosing a battle, can be pleasurable for some players (anarchists) while not for others (democrats) \cite{Hoogen2012}. 

However, when addressing the subject of video games entertainment, it is always assumed that the player has complete control over the character, regardless of whether it is a single player game or a competitive/cooperative game. TPP differs from those cases in the fact that everyone controlled the same character. As a consequence, enjoyment is no longer related to what a player, as a single individual, has done but rather to what they, as a group, have achieved. From the social identity approach perspective this can be described as a shift from the personal identity to the group identity. This shift would increase conformity to the norms associated to each group but as the groups were unstructured their norms would be inferred from the actions taken by the rest of the group \cite{reicher2001psychology}. New group members would then perform the actions they saw appropriate for them as members of the group, even if they might be seen as antinormative from an outside perspective \cite{spears2015}. This can be clearly seen in the behavior of the anarchists. Indeed, every time the game entered in democracy mode, anarchists started to send $start9$ as a form of protest, hijacking the democracy. Interestingly, this kept happening even though most of the players who were in the original protest did not play anymore (see figure S8). Thus, newcomers adopted the identity of the group even if they had not participated in its conception. Even more, stalling the game might have been regarded as antisocial behavior from the own anarchists point of view when they were playing under anarchy rules, but when the game entered into democracy mode it suddenly turned into an acceptable behavior, something that is predicted by the theory.

\begin{figure}[t]
\centering
\includegraphics[width=16cm]{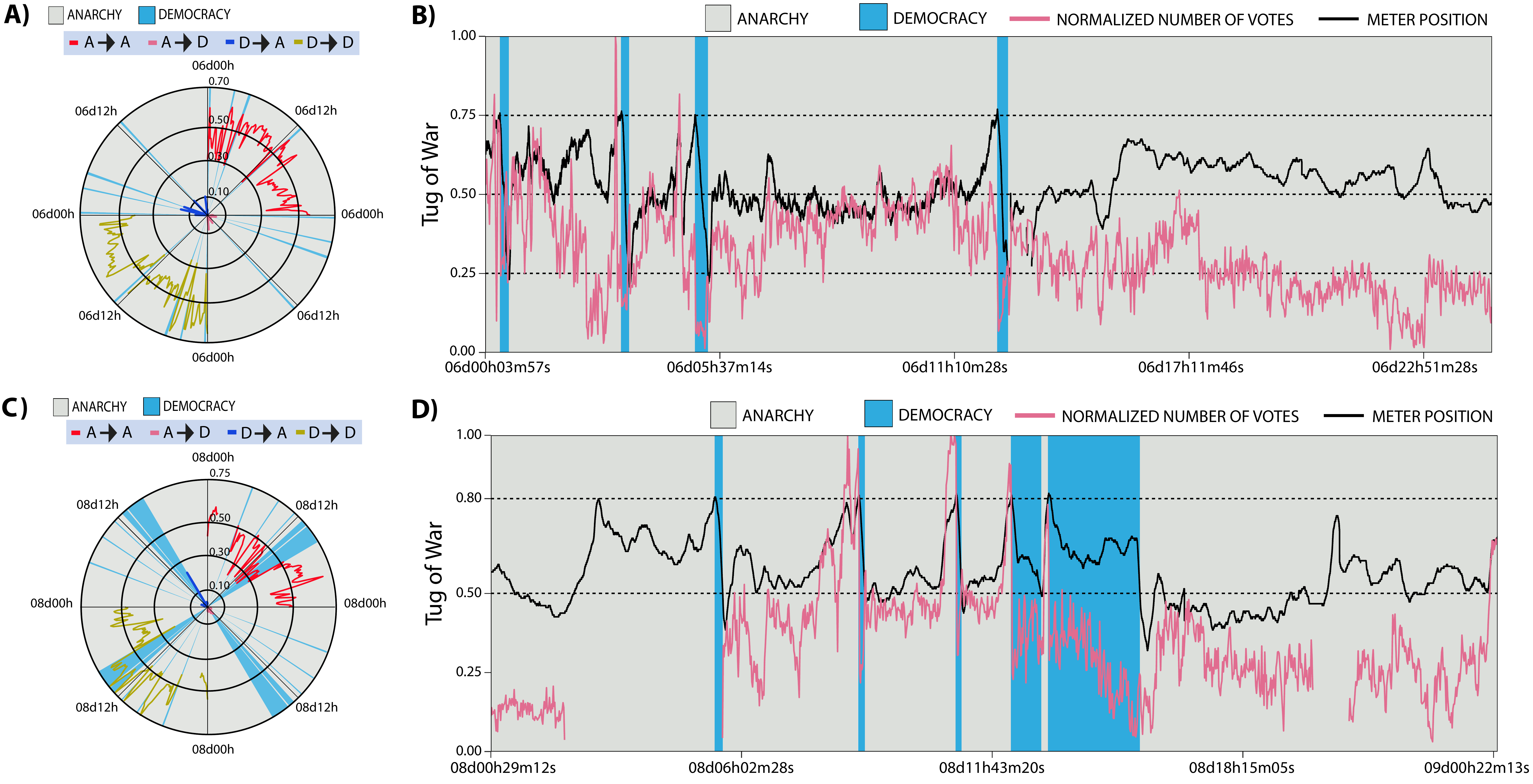}%
\caption{Politics of the crowd in days 6 (top) and 8 (bottom). In every plot the gray color represents when the game was played under anarchy rules and the blue color when it was played under democracy rules. The polar plots represent the evolution of the fraction of votes corresponding to anarchy/democracy while distinguishing if the user previously voted for anarchy or democracy: first quadrant, votes for anarchy coming from users who previously voted for anarchy ($A\rightarrow A$); second quadrant, votes for democracy coming from anarchy ($A\rightarrow D$); third quadrant, votes for democracy coming from democracy ($D\rightarrow D$); fourth quadrant, votes for anarchy coming from democracy ($D\rightarrow A$). In the other plots we show the evolution of the total number of votes for anarchy or democracy as a function of time normalized by its maximum value (green) as well as the position of the tug-of-war meter (black). When the meter goes above $0.75$ the system enters into democracy mode (blue) until it reaches $0.25$ (this threshold was later changed to $0.50$ ) when it enters into anarchy mode (gray) again.
}%
\label{fig:democracy}%
\end{figure}

To further explore the dynamics of these two groups, we next compare two different days: day 6 and day 8. Day 6 was the second day after the introduction of the anarchy/democracy dynamics and there were not any extremely difficult puzzles or similar areas where democracy might have been needed. On the other hand, day 8 was the day when the crowd arrived to the safari zone, which clearly needed the democracy mode, since the available number of steps in this area is limited (finite). The first thing we note is that, contrary to what we observed in section \ref{subsection:ledge}, in this case the commands coming from low activity users are not equivalent to the ones coming from high activity users. As a matter of fact, all of them can be removed and the results would still be the same (see figures S9 and S10). Our results are summarized in figure \ref{fig:democracy}.

One of the most characteristic features of groups is their polarization \cite{sunstein1999} \cite{conover2011political}. The problem in this particular case is that as people were leaving the game while others were constantly coming it is not straightforward to measure polarization. The fact that the number of votes for democracy could increase at a given moment did not mean that anarchists changed their opinion, it could be that new users were voting for democracy or simply that players who voted for anarchy stopped voting. Then, to properly measure polarization we consider 4 possible states for each user. They are defined by both the current vote of the player and the immediately previous one (note that we have removed players who only voted once): $A\rightarrow A$, first anarchy then anarchy; $A \rightarrow D$, first anarchy then democracy; $D \rightarrow D$, first democracy then democracy; $D \rightarrow A$, first democracy then anarchy. As we can see in figures \ref{fig:democracy}A and \ref{fig:democracy}C the communities are very polarized, with very few individuals changing their votes. The fraction of users changing from anarchy to democracy is always lower than 5\%, which indicates that anarchists form a very closed group. Similarly, the fraction of users changing from democracy to anarchy is also very low, although there are clear bursts when the crowd exits the democracy mode. This reflects that those who changed their vote from anarchy to democracy do so to achieve a particular goal, as going through a mace, and once they achieve the target, they instantly lose interest in democracy.

With such degree of polarization the next question is how was it possible for the crowd to change from one mode to the other. In figure \ref{fig:democracy}B we can see that every time the meter gets above the democracy threshold it is preceded by an increase in the total number of votes. Then, once under democracy mode the total number of votes decays very fast. Finally, there is another increment before entering again into anarchy mode. Thus, it is clear that every time democrats were able to enter into their mode they stopped voting and started playing. This let anarchists regain control even though they were less users, leading to a sharp decay of the tug of war meter. Once they exited democracy mode, democrats started to vote again to try to set the game back into democracy mode. In figure \ref{fig:democracy}D we can see initially a similar behavior in the short periods when democracy was installed. However, there is a wider area were the crowd accepted the democracy, this marks the safari zone mentioned previously. Interestingly, we can see how democrats learned how to keep their mode active. Initially there was the same drop on users voting and on the position of the meter seen in the other attempts. This forced democrats to keep voting instead of playing, which allowed them to retain control for longer. Few minutes later the number of votes decays again but in this case the position of the meter is barely modified probably due to anarchists accepting that they needed democracy mode to finish this part. Even though they might have implicitly accepted democracy, it is interesting to see that the transitions $A \rightarrow D$ are minimum (figure \ref{fig:democracy}C). Finally once the mission for which the democracy mode was achieved, finishing the safari zone, there is a sharp increment in the fraction of transitions $D \rightarrow A$.

\section{Conclusions}

In this work we have analyzed a crowd based event where nearly 1 million users played a game with the exact same character. Remarkably, the event was not only highly successful in terms of participants but also in length, lasting for over two weeks. Despite its uniqueness and its out of the lab setting, we were able to obtain all the necessary data to analyze the formation and evolution of the event.

We have shown that the success of the event might be somehow counterintuitively related to the frustration it generated. One of the particularly frustrating areas was the ledge, a part of the game that can be completed in a few minutes but that took over 15 hours to complete. We have seen that in this area the behavior of low and high activity users is quite similar, even though they might have been unaware of it. Besides, we have built a model to explain how the crowd was able to finally exit this part and shown how a minority - either in the form of griefers, smart users or simply na\"ive individuals - can lead the crowd to a successful outcome, even in the lack of consensus.

We have then analyzed the effects that the introduction of a voting system had in the crowd. We have seen how the crowd was split into two groups. We have been able to explain the behavior of these groups using the social identity approach and seen how norms could last within groups longer than their own members. Finally we have studied the polarization of the crowd and addressed how it was possible to change from one mode to another even with extremely high polarization.

To sum up, we have seen that a huge online crowd can be studied using tools developed for completely different purposes. We hope that these results will shed some light on the behavior of online crowds and their role in collective action.

\begin{acknowledgements}
We thank K. C. Bathina, J. Bollen, J. P. Gleeson, and M. Quayle for helpful comments and suggestions. A.A. acknowledges the support of the FPI doctoral fellowship from MINECO and its mobility scheme. Y.M. acknowledges partial support from the Government of Arag\'on, Spain through a grant to the group FENOL, and by MINECO and FEDER funds (grant FIS2017-87519-P). The funders had no role in study design, data collection and analysis, or preparation of the manuscript.
\end{acknowledgements}

\bibliography{references}

\FloatBarrier
\newpage

\begin{center}
\textbf{\large Supplemental Materials: Collective social behavior in a crowd controlled game}
\end{center}
\setcounter{equation}{0}
\setcounter{figure}{0}
\setcounter{table}{0}
\setcounter{page}{1}
\makeatletter
\renewcommand{\theequation}{S\arabic{equation}}
\renewcommand{\thefigure}{S\arabic{figure}}
\renewcommand{\bibnumfmt}[1]{[S#1]}
\renewcommand{\citenumfont}[1]{S#1}

\begin{figure}[h]
\begin{center}
\includegraphics[width=\linewidth]{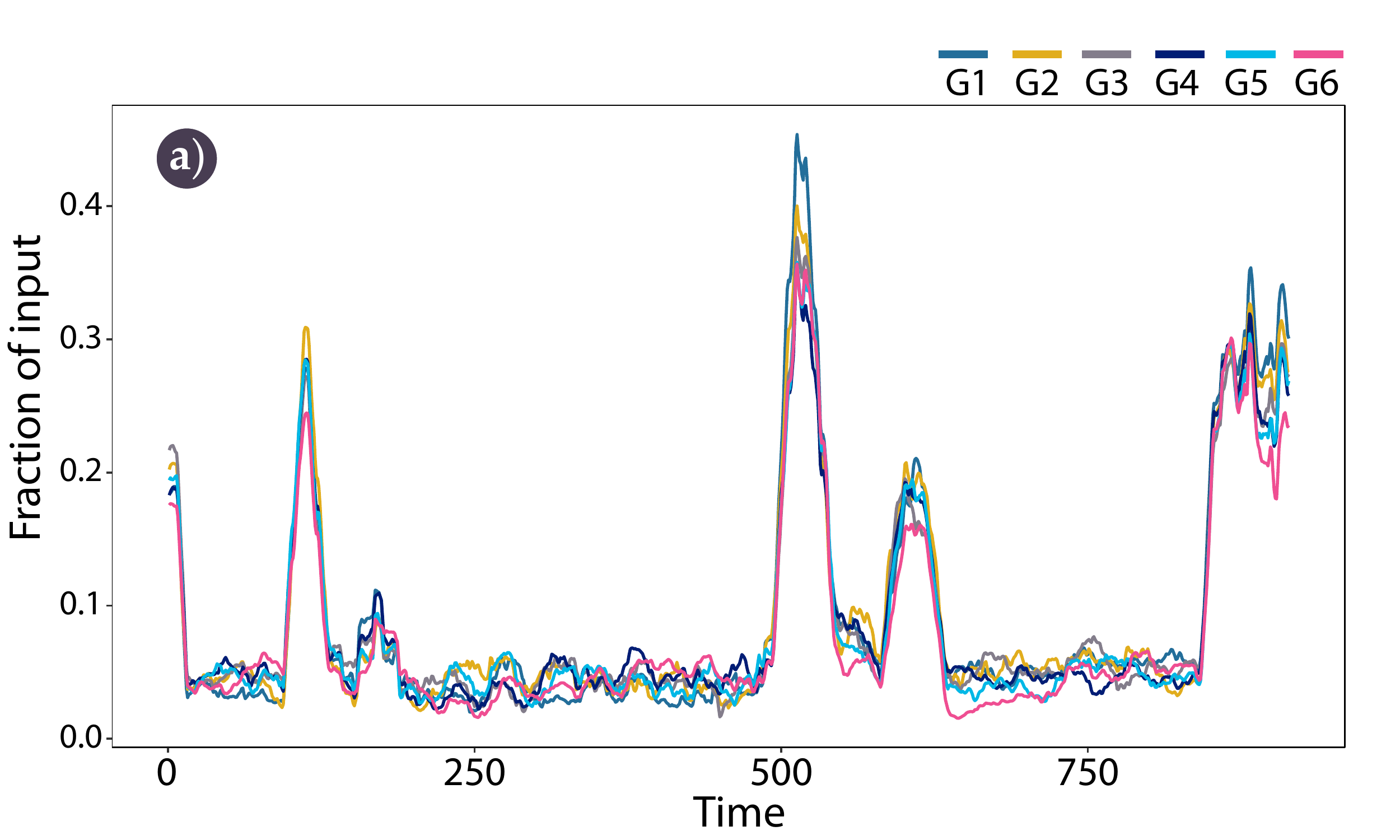}%
\end{center}
\caption{Time series of the fraction of \emph{a} in the input of each group.}%
\label{fig:supp_t_a}%
\end{figure}

\begin{figure}%
\begin{center}
\includegraphics[width=\linewidth]{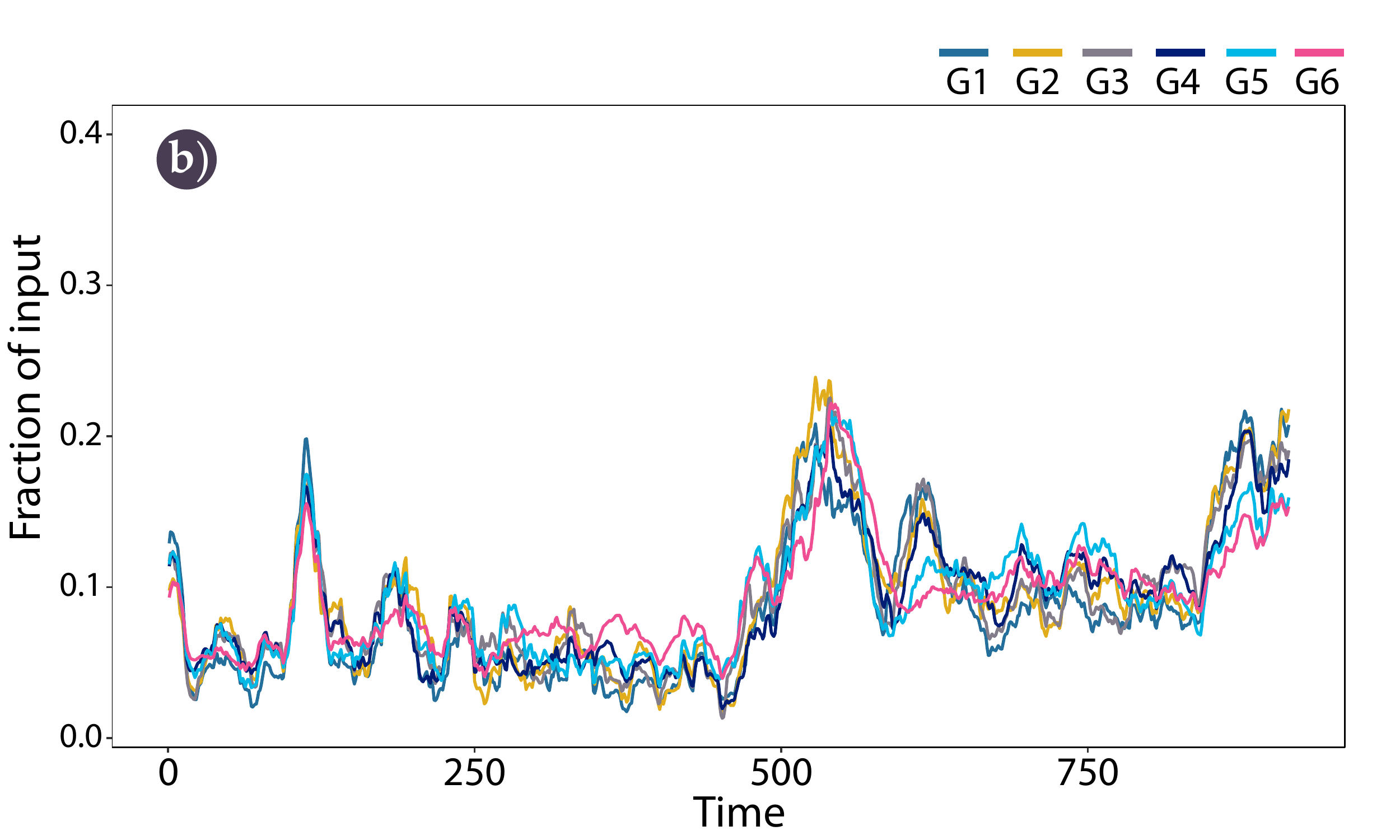}%
\end{center}
\caption{Time series of the fraction of \emph{b} in the input of each group.}%
\label{fig:supp_t_b}%
\end{figure}

\begin{figure}%
\begin{center}
\includegraphics[width=\linewidth]{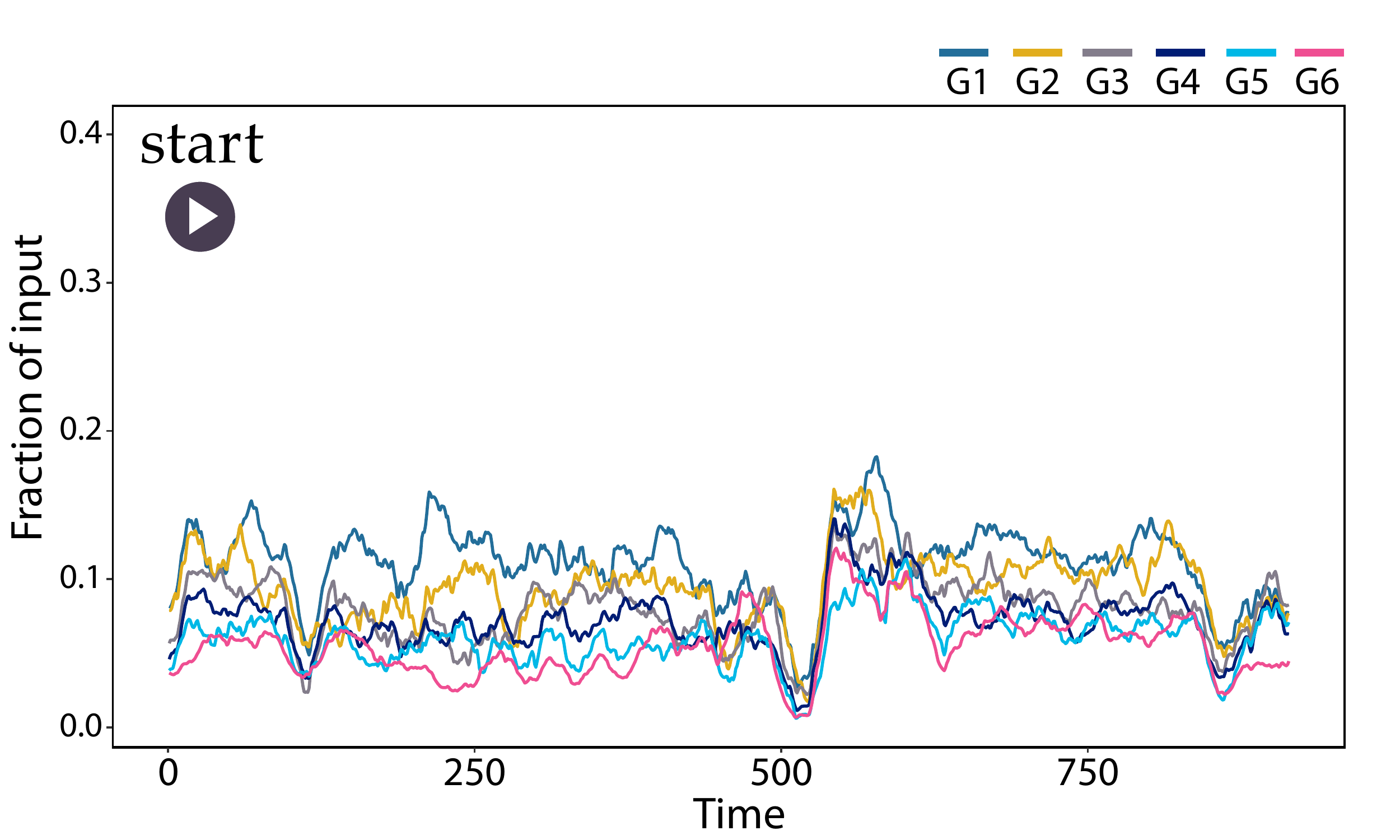}%
\end{center}
\caption{Time series of the fraction of \emph{start} in the input of each group.}%
\label{fig:supp_t_start}%
\end{figure}

\begin{figure}%
\begin{center}
\includegraphics[width=\linewidth]{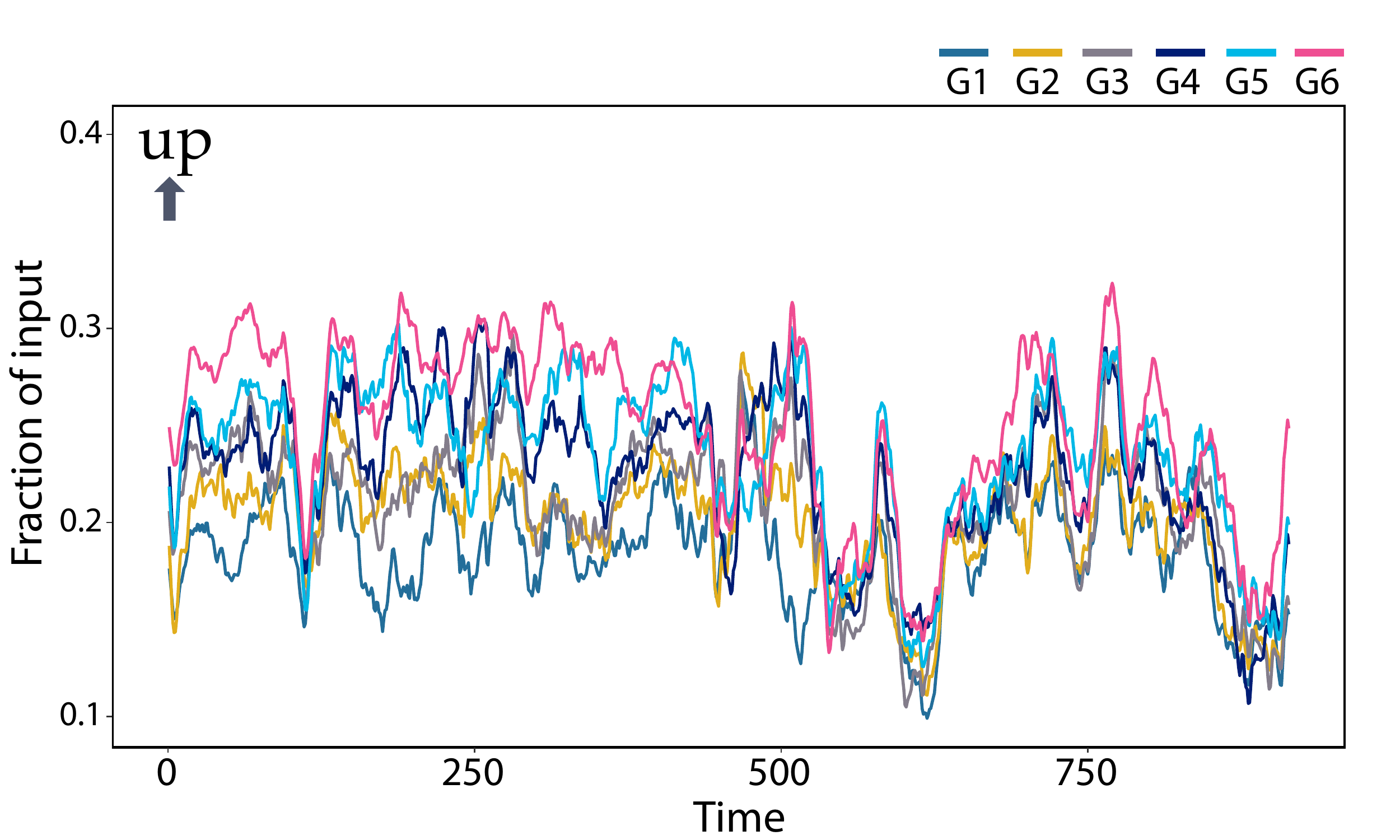}%
\end{center}
\caption{Time series of the fraction of \emph{up} in the input of each group.}%
\label{fig:supp_t_up}%
\end{figure}

\begin{figure}%
\begin{center}
\includegraphics[width=\linewidth]{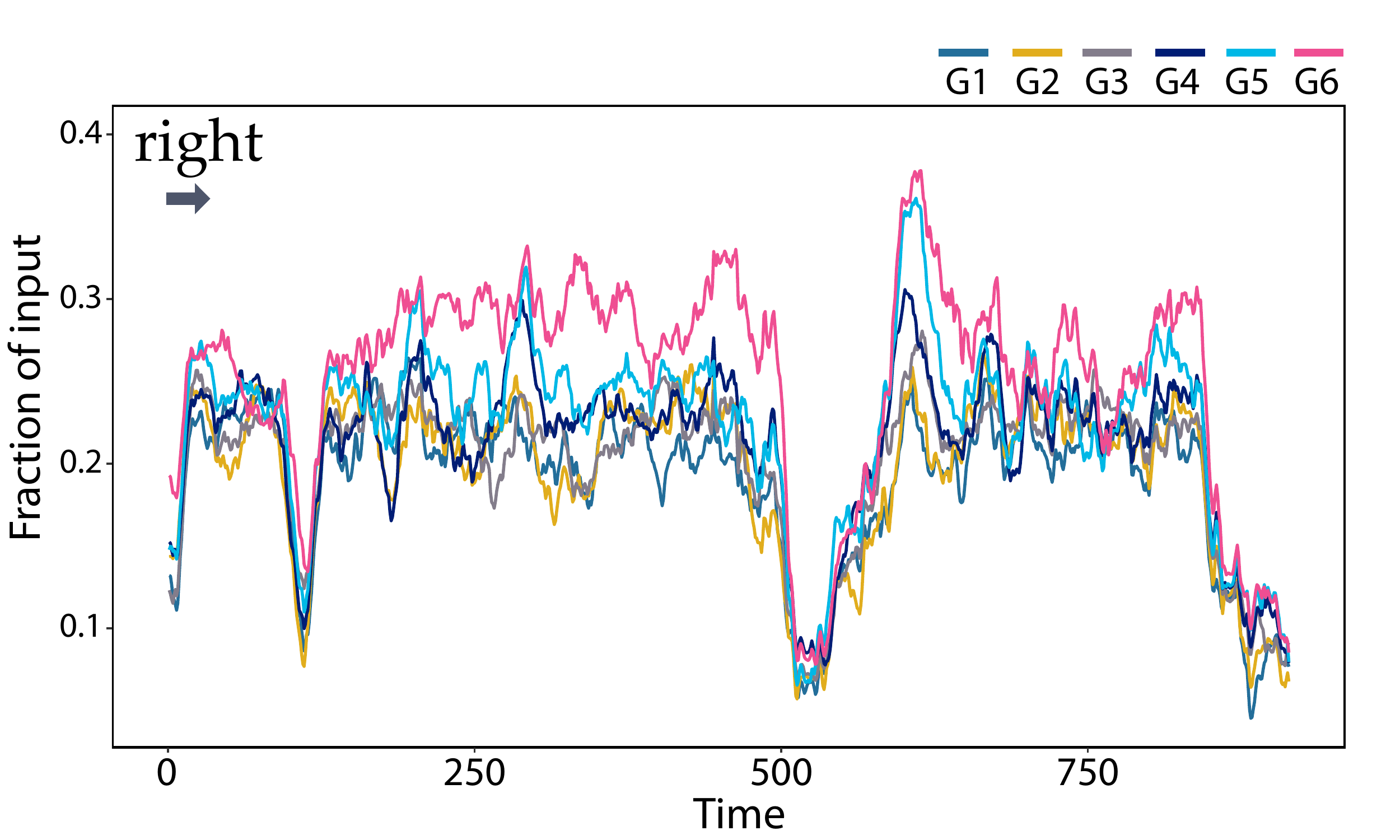}%
\end{center}
\caption{Time series of the fraction of \emph{right} in the input of each group.}%
\label{fig:supp_t_right}%
\end{figure}

\begin{figure}%
\begin{center}
\includegraphics[width=\linewidth]{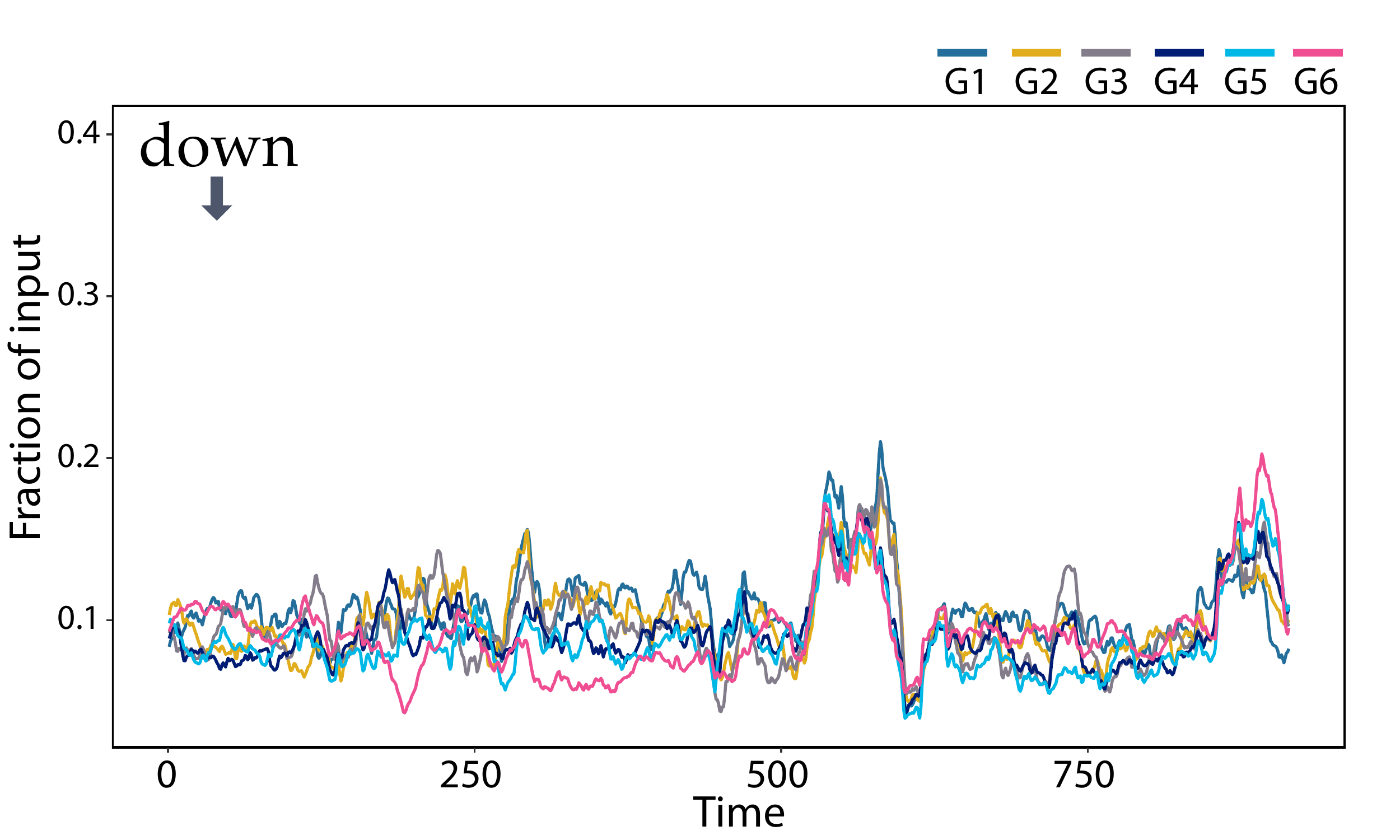}%
\end{center}
\caption{Time series of the fraction of \emph{down} in the input of each group.}%
\label{fig:supp_t_down}%
\end{figure}

\begin{figure}%
\begin{center}
\includegraphics[width=\linewidth]{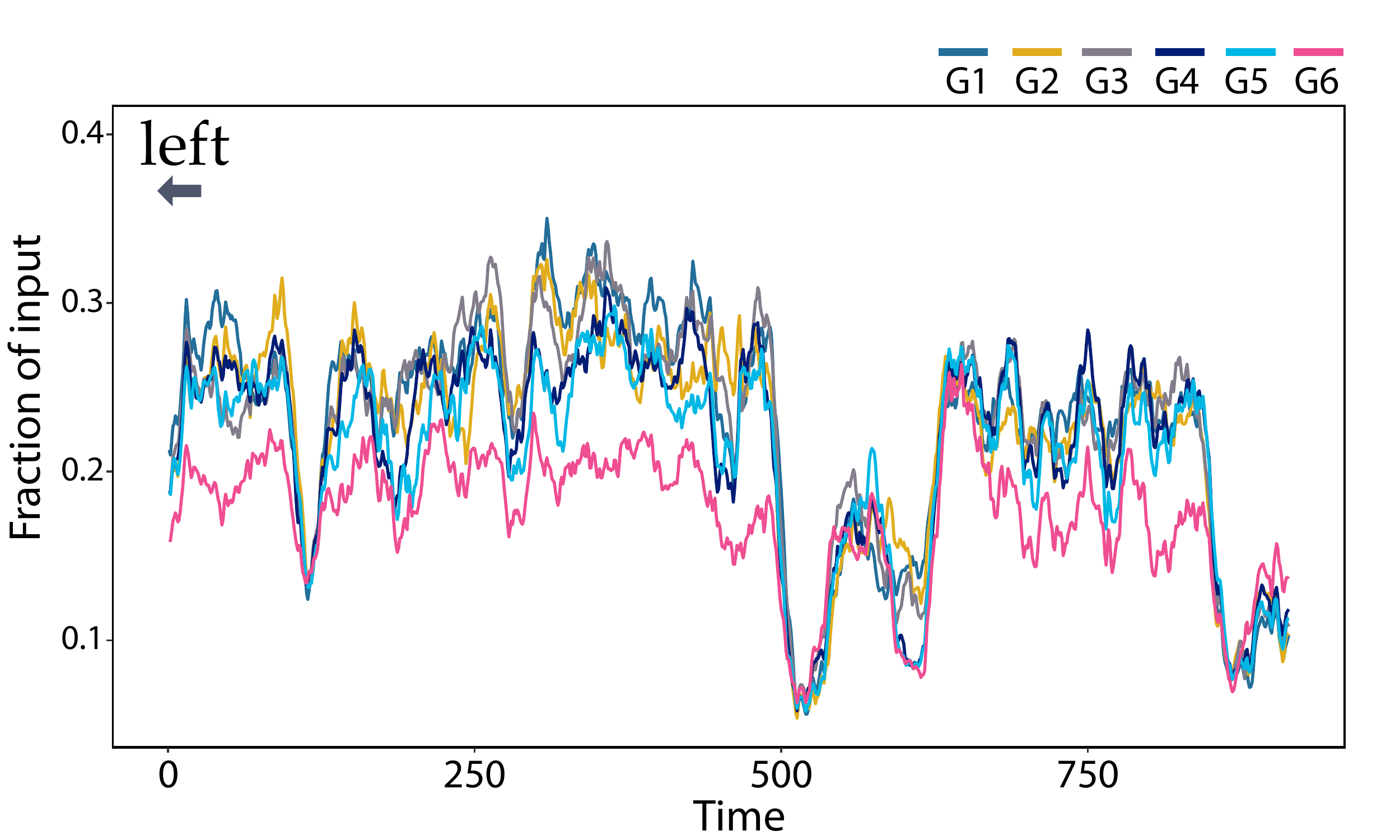}%
\end{center}
\caption{Time series of the fraction of \emph{left} in the input of each group.}%
\label{fig:supp_t_left}%
\end{figure}

\begin{figure}%
\begin{center}
\includegraphics[width=\linewidth]{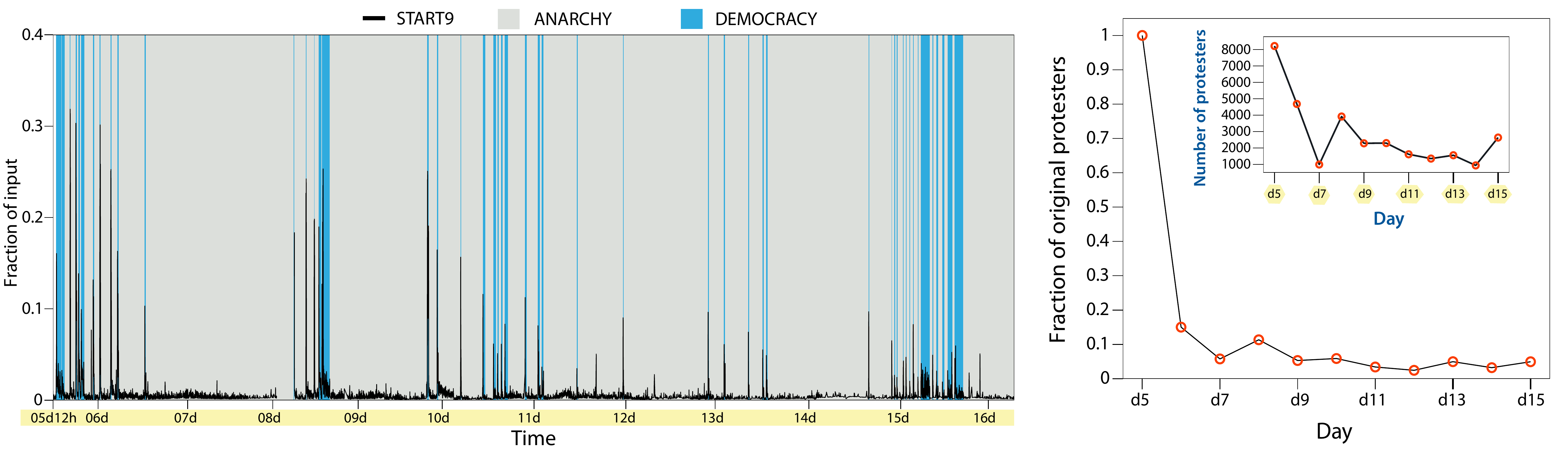}%
\end{center}
\caption{$start9$ protests throughout the game. Left: fraction of input corresponding to the $start9$ command. Right: fraction of users sending $start9$ who where in the original $start9$ riot (inset, total number of protesters each day). There were $start9$ protests 10 days after the first one even though very few of the original posters were still playing.}%
\label{fig:supp_start9}%
\end{figure}

\begin{figure}%
\begin{center}
\includegraphics[width=\linewidth]{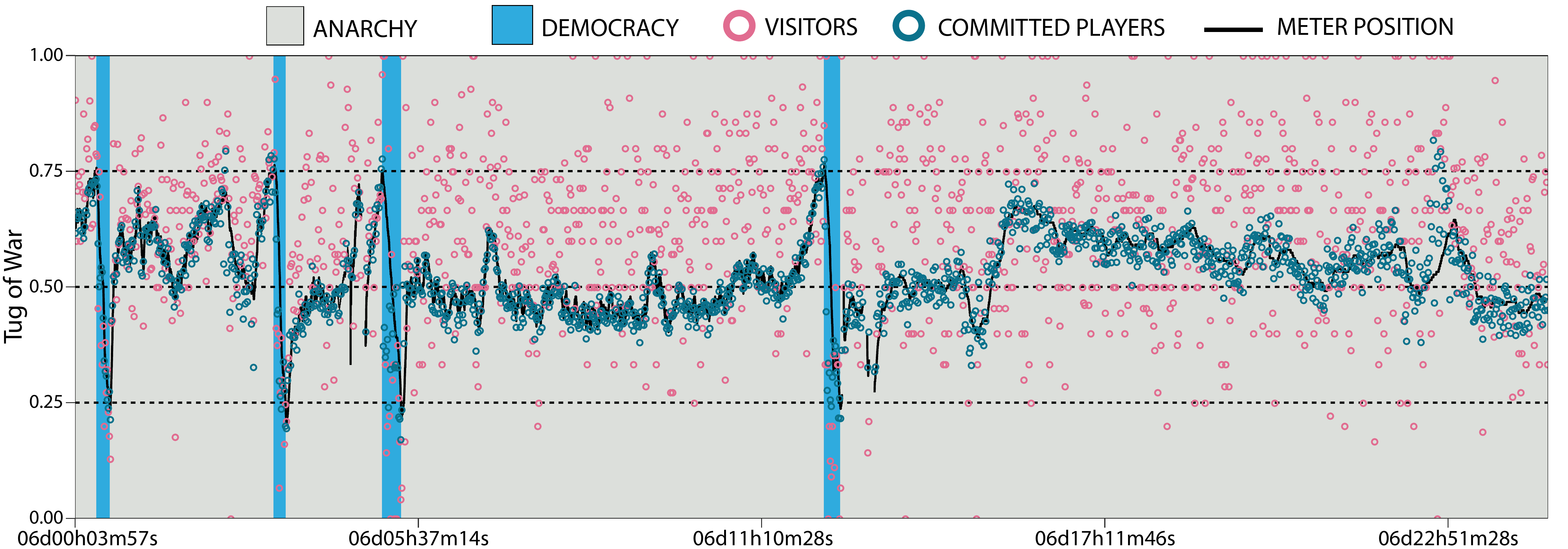}%
\end{center}
\caption{Meter position of the political tug of war if only votes from committed players (those who sent at least 2 votes) are taken into account (blue) and if only votes from visitors (those players who only participated once in the voting) are taken into account (pink). }%
\label{fig:supp_votes_comp1}%
\end{figure}

\begin{figure}%
\begin{center}
\includegraphics[width=\linewidth]{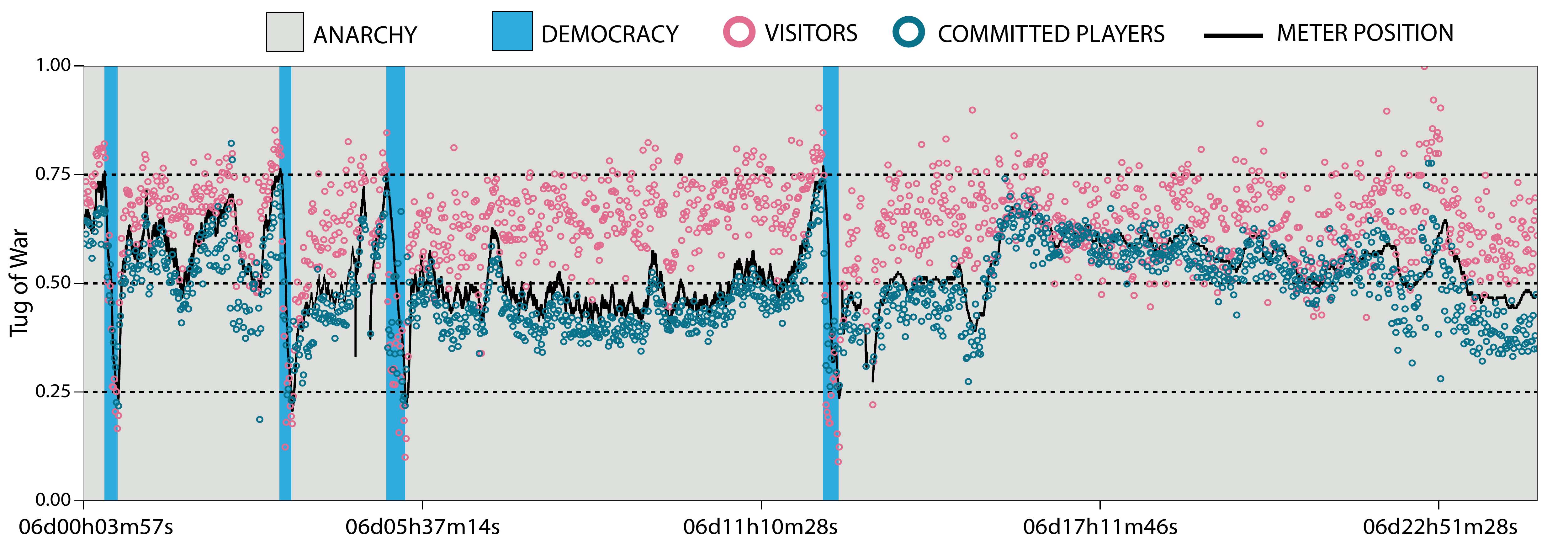}%
\end{center}
\caption{Meter position of the political tug of war if only votes from committed players (those who sent at least 10 votes) are taken into account (blue) and if only votes from visitors (those players who sent less than 10 votes) are taken into account (pink). }%
\label{fig:supp_votes_comp2}%
\end{figure}

\end{document}